\documentclass{article}
\usepackage{epsf,latexsym,amssymb}
\usepackage[mathscr]{eucal} 

\newcommand{\fractext}[2]{#1/#2}

\newcommand{\R}{{\mathbb R}}  

\newcommand{\text}[1]{\hbox{\rm \ #1\ \/}}

\newtheorem{theorem}{Theorem}
\newtheorem{itlemma}{Lemma}[section] 
\newtheorem{itproposition}[itlemma]{Proposition}
\newtheorem{itcorollary}[itlemma]{Corollary}
\newtheorem{itremark}[itlemma]{Remark}
\newtheorem{itdefinition}[itlemma]{Definition}
\newtheorem{itexample}[itlemma]{Example}

\newenvironment{lemma}{\begin{itlemma}\rm}{\end{itlemma}} 
\newenvironment{remark}{\begin{itremark}\rm}{\end{itremark}} 
\newenvironment{corollary}{\begin{itcorollary}\rm}{\end{itcorollary}}
\newenvironment{proposition}{\begin{itproposition}\rm}{\end{itproposition}}
\newenvironment{definition}{\begin{itdefinition}\rm}{\end{itdefinition}}
\newenvironment{example}{\begin{itexample}\rm}{\end{itexample}}

\newcommand{\be}[1]{\begin{equation}\label{#1}}
\newcommand{\ee}{\end{equation}}
\newcommand{\bl}[1]{\begin{lemma}\label{#1}}
\newcommand{\ble}[1]{\begin{lemmaex}\label{#1}}
\newcommand{\br}[1]{\begin{remark}\label{#1}}
\newcommand{\bt}[1]{\begin{theorem}\label{#1}}
\newcommand{\bd}[1]{\begin{definition}\label{#1}}
\newcommand{\bp}[1]{\begin{proposition}\label{#1}}
\newcommand{\bc}[1]{\begin{corollary}\label{#1}}
\newcommand{\bfact}[1]{\begin{fact}\label{#1}}
\newcommand{\ber}[1]{\begin{exercise}\label{#1}}
\newcommand{\bex}[1]{\begin{example}\label{#1}}
\newcommand{\bem}[1]{\begin{example}\label{#1}}  
\newcommand{\ec}{\mybox\end{corollary}}
\newcommand{\efact}{\mybox\end{fact}}
\newcommand{\eer}{\mybox\end{exercise}}
\newcommand{\eex}{\mybox\end{example}}
\newcommand{\eem}{\mybox\end{example}}
\newcommand{\el}{\mybox\end{lemma}}
\newcommand{\ele}{\mybox\end{lemmaex}}
\newcommand{\er}{\mybox\end{remark}}
\newcommand{\et}{\qed\end{theorem}}
\newcommand{\ed}{\mybox\end{definition}}
\newcommand{\ep}{\mybox\end{proposition}}
\newcommand{\epr}{\end{proof}}
\newcommand{\bpr}{\begin{proof}}

\newcommand{\ecs}{\end{corollary}}
\newcommand{\eers}{\end{exercise}}
\newcommand{\eexs}{\end{example}}
\newcommand{\eems}{\end{example}}
\newcommand{\els}{\end{lemma}}
\newcommand{\eles}{\end{lemmaex}}
\newcommand{\ers}{\end{remark}}
\newcommand{\ets}{\end{theorem}}
\newcommand{\eds}{\end{definition}}
\newcommand{\eps}{\end{proposition}}
\newcommand{\halmos}{\rule{1ex}{1.4ex}}
\newcommand{\qed}{\hfill \halmos} 
\newcommand{\mybox}{\hfill $\Box$} 
\newcommand{\beq}{\begin{eqnarray}}
\newcommand{\eeq}{\end{eqnarray}}
\newcommand{\beqn}{\begin{eqnarray*}}
\newcommand{\eeqn}{\end{eqnarray*}}
\newcommand{\bi}{\begin{itemize}}
\newcommand{\ei}{\end{itemize}}
\newcommand{\ben}{\begin{enumerate}}
\newcommand{\een}{\end{enumerate}}
\newcommand{\st}{\, | \,}

\newcommand{\col}{{\rm col}\,}
\newenvironment{proof}{\noindent {\em Proof}.\ }{\hspace*{\fill}$\halmos$\medskip}

\newcommand{\bs}{\begin{split}}
\newcommand{\es}{\end{split}}

\newcommand{\comment}[1]{}
\newcommand{\edo}{\end{document}}

\newcommand{\pbar}{\bar p}
\newcommand{\xbar}{\bar x}
\newcommand{\eqmul}[2]{\begin{equation}\label{#1}\begin{array}{rcl}
#2\end{array}\end{equation}}
\newcommand{\xo}{x^{\scriptscriptstyle 0}}
\newcommand{\tbar}{\bar t}
\newcommand{\xbaro}{{\bar x}^{\scriptscriptstyle 0}}
\newcommand{\perturbj}{d_jp}

\title{Determination of Functional Network Structure\\ from
Local Parameter Dependence Data}

\author{Boris N. Kholodenko\\
Department of Pathology, Anatomy and Cell Biology\\
Thomas Jefferson University, PA\\
{\tt Boris.Kholodenko@mail.tju.edu} \and
Eduardo D. Sontag
\\
Dept. of Mathematics\\
Rutgers University, NJ, USA\\
{\tt sontag@hilbert.rutgers.edu}}

\date{}

\begin{document}
\maketitle

\begin{abstract}
\noindent
In many applications, such as those arising from the field of cellular
networks, it is often desired to determine the interaction (graph)
structure of a set of differential equations, using as data measured
sensitivities.  This note proposes an approach to this problem.
\end{abstract}

\section{Introduction}

Suppose given a system of differential equations
\eqmul{system}{
\dot x_1 &=& f_1(x_1,\ldots ,x_n,p_1,\ldots ,p_m) \\
\dot x_2 &=& f_2(x_1,\ldots ,x_n,p_1,\ldots ,p_m) \\
        &\vdots&\\
\dot x_n     &=& f_n(x_1,\ldots ,x_n,p_1,\ldots ,p_m) \,,
}
where the vector of state variables $x(t)=(x_1(t),\ldots ,x_n(t))$ evolves in
some open subset $X\subseteq \R^n$ and the vector of parameters $p=(p_1,\ldots ,p_m)$ can
be chosen from an open subset ${\cal P}\subseteq \R^m$, for some $n$ and $m$.
In cellular networks, for instance, the state variables $x_i$ might
represent the concentrations of certain proteins, mRNA, etc., at different time
instants, and the parameters $p_i$ might represent the concentration levels
of certain enzymes which are maintained at a constant value during a
particular experiment (see, e.g.,~\cite{k1,k2}).

In many fields of application, it is often the case that the equations
defining the system (that is, the form of the functions $f_i$ describing the
vector field) are unknown, even in general form, but one wishes nonetheless
to determine the interaction graph of a system~(\ref{system}), that is to
say, to know which variables directly influence which variables, as well as
the relative strengths of these interactions (\cite{k3}).  A more limited goal
might be to determine if a certain feedback loop is present in the system
(cf.~\cite{k5}).

To help in this task, experimental data is usually available, measuring
solutions of system~(\ref{system}) for various initial states and parameter
settings.  A special case, the one treated in this note, is that in which
experimental data provides us with the location of steady states associated to
parameter values near a given set of parameters $\pbar$.
We show in this note how, starting from such data, and under an assumption
that seems natural, it is indeed possible to determine this interaction graph
and relative strengths of interactions.
We then extend the method to non-steady state measurements.
A more global analysis of the problem is also possible, and will be presented
in a follow-up report.

\section{Problem Formulation}

The steady states of system~(\ref{system}) associated to a given parameter
vector $p=(p_1,\ldots ,p_m)$ are the solutions $x=(x_1,\ldots ,x_n)$ of
the set of $n$ simultaneous algebraic equations
\eqmul{system-alg}{
f_1(x_1,\ldots ,x_n,p_1,\ldots ,p_m) &=&0\\
f_2(x_1,\ldots ,x_n,p_1,\ldots ,p_m)  &=&0\\
        &\vdots&\\
f_n(x_1,\ldots ,x_n,p_1,\ldots ,p_m)  &=&0\,.
}
We will assume that there is a function $\xi :{\cal P}\rightarrow {\cal X}$ which assigns, to each
parameter vector $p\in {\cal P}$, a steady state $\xi (p)$ of~(\ref{system}), that is to
say, one has that $f_i(\xi (p),p)=0$ for all $i=1,\ldots ,n$ and all $p\in {\cal P}$.
We suppose that the functions $f_i$ as well as $\xi $ are continuously
differentiable.
A particular parameter vector $\pbar$ is also given, and the problem will be
formulated in terms of the behavior of steady states near $\xbar=\xi (\pbar)$.

\subsubsection*{Experimental Data}

It will be assumed that an $n\times m$ matrix $\Sigma =(\sigma _{kj})$ is given, representing
the ``sensitivities''
\be{sensitivities}
\sigma _{kj} = \frac{\partial \xi _k}{\partial p_j} (\pbar)
\ee
for each $k=1,\ldots ,n$ and each $j=1,\ldots ,m$.
This matrix of partial derivatives may be estimated numerically from the
values of $\xi (p)$ on a neighborhood of the chosen parameter $\pbar$.

\subsubsection*{Desired Information}

Consider the $n\times n$ matrix $A=(a_{ij})$ defined by:
\[
a_{ij} = \frac{\partial f_i}{\partial x_j} (\xbar,\pbar)
\]
for every $i,j=1,\ldots ,,n$.

Ideally, one would want to find the matrix $A$, since this matrix completely
describes the influence of each variable $x_j$ upon the rate of change of each
other variable $x_i$.
Unfortunately, such an objective is impossible to achieve from the local
steady-state data $\Sigma $, or even from the knowledge of the complete (global)
mapping $\xi $.  This is because the same mapping $\xi $ also solves the set of
equations
\eqmul{system-alg-mod}{
\lambda _1 f_1(x_1,\ldots ,x_n,p_1,\ldots ,p_m) &=&0\\
\lambda _2 f_2(x_1,\ldots ,x_n,p_1,\ldots ,p_m)  &=&0\\
        &\vdots&\\
\lambda _n f_n(x_1,\ldots ,x_n,p_1,\ldots ,p_m)  &=&0\,,
}
for any constants $\lambda _i$, but, on the other hand, multiplication of $f_i$ by
$\lambda _i$ results in $a_{ij}=\lambda _ia_{ij}$.  In other words, the best that one
could hope is for the data $\Sigma $ to determine the rows
\[
A_i = (a_{i1},\ldots ,a_{in}) \,,\;\; i=1,\ldots ,n
\]
of $A$ only up to scalar multiples.

Thus, a more realistic objective is to attempt to identify the rows $A_i$
{\em up to a scalar multiple\/} only.
For example, if we assume that $a_{ii}\not= 0$ for each $i$ (a realistic
assumption when stable systems are being interconnected),
this amounts to finding the ratios $a_{ij}/a_{ii}$ for each $i\not= j$.

\subsubsection*{Assumptions}

We will make two assumptions which will suffice for us to solve the problem of
determining the rows $A_i$ of $A$ up to scalar multiples.  The first
assumption is a strong but reasonable structural one, while the second
represents a weak algebraic nondegeneracy condition.

We will suppose known, for each $i\in \{1,\ldots ,n\}$, a subset $S_i$ of the index
set $\{1,\ldots ,m\}$ so that the following property holds:
\be{indep-assumption}
(\forall\, j\in S_i) \;\;\; \frac{\partial f_i}{\partial p_j} (\xbar,\pbar) = 0
\ee
which is in turn implied by the structural condition:
\[
(\forall\, j\in S_i) \;\;\;  f_i \text{does not depend upon} p_j \,.
\]
This prior information about the system structure is far less restrictive than
it might appear at first sight.  Indeed, it is usually the case that
``compartmental'' information is available, for instance telling us that the
concentration of a certain enzyme has no direct influence on an unrelated
biochemical reaction, that an extracellular signaling molecule does not affect
directly a cytoplasmatic reaction, and so forth.

The second assumption is as follows.
For each $j\in \{1,\ldots ,m\}$, we introduce the vector
\be{Sj-def}
\Sigma _j = \col(\sigma _{1j},\ldots ,\sigma _{nj}) =
\pmatrix{
  \frac{\partial \xi _1}{\partial p_j} (\pbar)\cr\vdots\cr
\frac{\partial \xi _n}{\partial p_j}(\pbar)}
\ee
representing the $j$th column of the matrix $\Sigma $, and we consider, for
each $i\in \{1,\ldots ,n\}$ the linear subspace $H_i$ of $\R^n$ spanned by the
vectors
\[
\{\Sigma _j \st j\in S_i\}.
\]
The assumption is:
\[
\dim H_i \,\geq \, n-1 \quad\forall\,i=1,\ldots ,n\,.
\]
Note that this amounts to saying that the dimension of $H_i$ is either $n$ or
$n-1$.  (Generically, we may expect this dimension to be $n-1$, because
the orthogonality relation to be shown below is the only algebraic constraint.)

\section{Solution}

With the above assumptions, the problem that we posed can be solved as follows.
We fix any index $i\in \{1,\ldots ,n\}$, and take partial derivatives in the
equation $f_i(x(p),p)=0$ with respect to the variable $p_j$, for each index
$j$ in the set $S_i$, and evaluate at $x=\xbar$ and $p=\pbar$:
\[
0\;=\; \frac{\partial}{\partial p_j}f_i(\xbar,\pbar) \;=\;
\sum_{k=1}^n \frac{\partial f_i}{\partial x_k} (\xbar,\pbar) \,
             \frac{\partial x_k}{\partial p_j} (\xbar)\,
+ \,\frac{\partial f_i}{\partial p_j} (\xbar,\pbar)
 \;=\;
A_i\cdot \Sigma _j
\]
where the second term vanishes by assumption~(\ref{indep-assumption}).

Since this happens for every $j\in S_i$,
we conclude that the vector $A_i$ is orthogonal to $H_i$.  As $H_i$ has
dimension $n$ or $n-1$, this determines $A_i$ up to a scalar multiple, which
is what we wanted to prove.  

Of course, it is trivial to actually compute $A_i$ from the data.
If $H_i$ has dimension $n$, then $A_i=0$; this is a degenerate case.
When the dimension is $n-1$, one simply picks a basis 
$\{\Sigma _{j_1},\ldots ,\Sigma _{j_{n-1}}\}$ of $H_i$, and any vector
$\Sigma _0$ linearly independent from the elements of this basis (a randomly chosen
vector has this property), and then solves
(for example) the nonsingular set of equations
$A_i\cdot \Sigma _0=1$, $A_i\cdot \Sigma _{j_\ell}=0$, $\ell=1,\ldots ,n-1$,
to find a nonzero $A_i$ (all possible $A_i$ are scalar multiples of this one).
Alternatively, provided that one knows that $a_{ii}\not= 0$, one may simply
normalize to $a_{ii}=-1$ and then determine the remaining entries of $A_i$
by solving a linear set of $n-1$ equations.
Observe also, that if it is known {\em a priori} that certain entries $a_{ij}$
vanish, then one may redefine the space $H_i$ to be spanned only by the
vectors listing the appropriate components of the sensitivities
${\partial \xi _i}/{\partial p_j}$'s, and a potentially much smaller number of
parameter perturbations may be required.

\section{Modular Approach}

It is also possible to apply our techniques in a ``modular'' context, in which
only the derivatives $\fractext{\partial f_i}{\partial x_j}$ with respect to
communicating intermediaries are calculated (\cite{k5}).  Let us briefly
explain this. 

We assume that the entire network consists of an interconnection of $n$
subsystems or ``modules'', each of which is described by a set of differential
equations such as:
\eqmul{system-modules}{
\dot x_{j}   &=& g_{0,j}(y_{1,j},\ldots ,y_{\ell,j},x_1,\ldots ,x_n,p_1,\ldots ,p_m) \\
\dot y_{1,j} &=& g_{1,j}(y_{1,j},\ldots ,y_{\ell,j},x_1,\ldots ,x_n,p_1,\ldots ,p_m) \\
\dot y_{2,j} &=& g_{2,j}(y_{1,j},\ldots ,y_{\ell,j},x_1,\ldots ,x_n,p_1,\ldots ,p_m) \\
        &\vdots&\\
\dot y_{\ell_j,j} &=& g_{\ell_j,j}(y_{1,j},\ldots ,y_{\ell,j},x_1,\ldots ,x_n,p_1,\ldots ,p_m)
\,, 
}
where the variables $x_j$ represent ``communicating'' or ``connecting''
intermediaries of module $j$ that transmit information to other modules,
whereas the variables $y_{1,j},\ldots ,y_{\ell,j}$ represent chemical species
that interact within module $j$.
The integer $\ell_j$, $j=1,\ldots ,n$ is in general different for each of the
$n$ modules and represents the number of chemical species in the $j$th module.

We will assume that, for each fixed module, the Jacobian of
$(g_1,\ldots ,g_{\ell_j,j})$ with respect to $y_1,\ldots ,y_{\ell_j}$, evaluated at
the steady state corresponding to $\pbar$ (assumed to exist, as before) is
nonsingular. 
The Implicit Mapping Theorem then implies that one may, in a neighborhood of
this steady state, solve 
\eqmul{eqs-system-modules}{
g_{0,j}(y_{1,j},\ldots ,y_{\ell,j},x_1,\ldots ,x_n,p_1,\ldots ,p_m) &=&0\\
g_{1,j}(y_{1,j},\ldots ,y_{\ell,j},x_1,\ldots ,x_n,p_1,\ldots ,p_m) &=&0\\
g_{2,j}(y_{1,j},\ldots ,y_{\ell,j},x_1,\ldots ,x_n,p_1,\ldots ,p_m) &=&0\\
        &\vdots&\\
g_{\ell_j,j}(y_{1,j},\ldots ,y_{\ell,j},x_1,\ldots ,x_n,p_1,\ldots ,p_m)&=&0
}
for the variables $x_j,y_1,\ldots ,y_{\ell_j}$ as a
function of $x_1,\ldots ,x_n,p_1,\ldots ,p_m$.
One concludes that, around this
steady state corresponding to $\pbar$, the functions $x_j$ satisfy
implicit equations of the form
\[
x_j = h_j(x_1,\ldots ,x_n,p_1,\ldots ,p_m)
\]
which we can rewrite in the form~(\ref{system-alg}), using
$f_j(x,p)=x_j-h_j(x,p)$.
The analysis then proceeds as before.
The generalization to the case of more than one communicating intermediate in
a module, namely a vector $(x_{j,1},\ldots ,x_{j,k_j})$, is obvious.

\section{Avoiding Derivatives}

The technique that was described assumes that we know the sensitivity matrix
$\Sigma $, which is obtained by evaluating the partial derivatives
$\fractext{\partial \xi _i}{\partial p_j}$ at the particular parameter value
$\pbar$.
Ordinarily, these derivatives would be estimated by finite differences.
For instance, suppose that one measures $\xbar=\xi (\pbar)$ as well as
$\xi (\pbar+\perturbj)$, where
\[
\perturbj = \col(0,\ldots ,0,dp_j,0,\ldots ,0)
\]
(entry in $j$th position)
and we view $dp_j$ as a ``small'' perturbation of the $j$th parameter.
Denoting
\be{djxi-def}
d_jx_i := \xi _i(\pbar+\perturbj) - \xbar_i
\ee
obviously one may estimate $\Sigma $ using the following approximation:
\[
\frac{\partial \xi _i}{\partial p_j} (\pbar) \;\approx\;
\frac{d_jx_i}{dp_j}\,.
\]
In order to calculate this ratio, both $d_jx_i$ and $dp_j$ must be known.

However, in certain experimental situations it may well be impossible to
estimate the values of $dp_j$.  This might appear to be contradictory, since
we are assuming that we perform experiments which change the values of $p$.
But one can easily envision an experimental setup in which a certain external
variable (in a cell biology situation, for instance, a growth factor) is
known to influence a certain parameter $p_j$.  Varying this external variable
therefore produces a perturbation in $p_j$, and hence an appropriate $d_jx$
which is measured, but $dp_j$ itself may be hard to measure.

It is rather surprising that we can still achieve our goal of estimating the
rows of $A$ (up to scalar multiples) even in the absence of information
about the $dp_j$'s!  To see intuitively why this is plausible, consider the
following argument.  Let us say that we have just a scalar parameter $p$ and a
scalar function $f(x_1,x_2)$ so that $f(\xi _1(p),\xi _2(p))\equiv 0$, and
that $\xbar=\xi (\pbar)=(0,0)$.
In a neighborhood of $p=\pbar$, we may assume that $f$ is linear,
so we have a linear relation (with unknown coefficients)
\[
a_1\xi _1(p) + a_2\xi _2(p) = 0\,.
\]
The method discussed so far would take derivatives at $p=\pbar$:
\[
a_1 \frac{d\xi _1}{dp}(\pbar) + a_2\frac{d\xi _2}{dp}(\pbar) = 0
\]
and thus (assuming that the derivative is not zero),
we know that the row $(a_1,a_2)$ must be a multiple of
$(-(\fractext{d\xi _2}{dp})(p),(\fractext{d\xi _1}{dp})(p))$.
A completely different argument (analogous to using a two-point as opposed to
a slope-point formula in order to find the equation of a line) would simply
take the original equation $a_1\xi _1(p) + a_2\xi _2(p) = 0$ (valid only
$p\approx\pbar$, since this was an approximation of $f$) and say that
the row $(a_1,a_2)$ must be a multiple of
$(-\xi _2(p),\xi _1(p))$, for any fixed $p\approx\pbar$.
There is no inconsistency between the two estimates, since they only differ
(approximately) by multiplication by the scalar $dp$:
\[
(-\xi _2(p),\xi _1(p))
\;\approx\;
(-(\fractext{d\xi _2}{dp})(p),(\fractext{d\xi _1}{dp})(p))\, dp
\]
and we only care about scalar multiples.
Let us now say this in general.

Since $f_i(\xi (\pbar+\perturbj),\pbar+\perturbj)-f(\xbar,\pbar)=0-0=0$,
we have, taking a Taylor expansion, that
\be{expand-f}
\frac{d}{dp_j}f_i(\xi (p),p)\Big|_{p=\pbar} dp_j \,+ \,o(dp_j)
 \;=\; 
\sum_{k=0}^n \frac{\partial f_i}{\partial x_k} (\xbar,\pbar)
\frac{\partial \xi _k}{\partial p_j} (\pbar)\,dp_j
\, + \, o(dp_j) \;=\; 0\,.
\ee
whenever $j\in S_i$.  Substituting
\be{expand-dx}
d_jx_k \;=\; \xi _k(\pbar+\perturbj) - \xbar_k
\;=\; \frac{\partial \xi _k}{\partial p_j} (\pbar) \,dp_j \,+\, o(dp_j)
\ee
into~(\ref{expand-f}), we conclude that
\[
\sum_{k=0}^n \frac{\partial f_i}{\partial x_k} (\xbar,\pbar)\,d_jx_k
\;=\; o(dp_j)
\;\text{provided that} j\in S_i\,.
\]
Since $dp_j\approx0$ and $d_jx_k=O(dp_j)$, this is an approximate
orthogonality relation, and we now make the approximation:
\be{approximate-orthogonality}
\sum_{k=0}^n \frac{\partial f_i}{\partial x_k} (\xbar,\pbar)\,d_jx_k
\;=\;0\;\text{provided that} j\in S_i\,.
\ee
In conclusion, and introducing the matrix $\Gamma =(\gamma _{kj})=(d_jx_k)$ instead of
$\Sigma $,
we have that $A_i\cdot \Gamma _j=0$ for all $j\in S_i$, where 
$\Gamma _j=\col(\gamma _{1j},\ldots ,\gamma _{nj})$.
Now we consider, for
each $i\in \{1,\ldots ,n\}$, the linear subspace $K_i$ of $\R^n$ spanned by the
vectors $\{\Gamma _j \st j\in S_i\}$ 
and assume that $\dim K_i\geq n-1$ for all $i$.
We conclude that the vector $A_i$ is orthogonal to $K_i$, and
this once again determines $A_i$ up to a scalar multiple.

\section{Non-Steady State Analysis}

Let us sketch here how one might extend our methodology to use non-steady
state data. 
In general, we denote by $\xi (t,\xo,p)$ the solution of~(\ref{system}) with
initial condition $\xo$, at time $t$ and using parameters $p$.
Let us suppose that we can measure the sensitivities~(\ref{sensitivities})
at some specific point in time, and for some specific solution
$\xi (\tbar,\xbaro,\pbar)$:
\be{sensitivities-tv}
\sigma _{kj} = \frac{\partial \xi _k}{\partial p_j} (\tbar,\xbaro,\pbar)
\ee
for each $k=1,\ldots ,n$ and each $j=1,\ldots ,m$, and we let $\Sigma =(\sigma _{kj})$.
We also need now the mixed second derivatives:
\[
\eta _{ij} = \frac{\partial^2 \xi _i}{\partial p_j\partial t} (\tbar,\xbaro,\pbar)
\]
and instead of $\Sigma _j=\col(\sigma _{1j},\ldots ,\sigma _{nj})$ as in~(\ref{Sj-def}), we
consider for each $i\in \{1,\ldots ,n\}$ and $j\in \{1,\ldots ,m\}$ the vector
\be{Sj-def-new}
\Sigma _{ij} = \col(\eta _{ij},\sigma _{1j},\ldots ,\sigma _{nj}).
\ee
We define, for each $i\in \{1,\ldots ,n\}$, $H_i$ as the linear subspace of
$\R^{n+1}$ spanned by the vectors
$
\{\Sigma _{ij} \st j\in S_i\}
$.
We let now
$a_{ij} = \frac{\partial f_i}{\partial x_j} (\tbar,\xbaro,\pbar)$.
Fixing any index $i\in \{1,\ldots ,n\}$, we take partial derivatives on both sides
of the differential equation
\[
\frac{\partial}{\partial t} \xi _i(t,\xo,p)
\;=\;
f_i(\xi (t,\xo,p),p)
\]
with respect to the
variable $p_j$, for each index $j$ in the set $S_i$, and evaluate at
$x=\xbar$, $t=\tbar$, and $p=\pbar$, to obtain:
\[
\eta _{ij}\;=\; 
\frac{\partial}{\partial p_j}f_i(\xbar,\pbar) \;=\;
\sum_{k=1}^n a_{ik} \sigma _{kj}
\]
from which we conclude that
$
[-1,A_i]\cdot \Sigma _j = 0
$
whenever $j\in S_i$, and hence that $[-1,A_i]$ is orthogonal to $H_i$.
With appropriate genericity conditions, this orthogonality, perhaps in
conjunction with conditions at other times $t$ or points $p$, will restrict
the possible vectors $A_i$ and more generally the interaction graph.  
(For example, if $\dim H_i=n$, then we have a unique solution.)
Derivatives with respect to parameter values can be replaces by differences,
just as in the steady state case.
We will discuss this further in a future contribution.

\edo